\documentstyle[12pt,psfig]{article}

\newcommand{\veps}{\varepsilon}
\newcommand{\beq}[1]{\begin{equation} \label{#1}}
\newcommand{\eeq}{\end{equation}}
\newcommand{\beqa}[1]{\begin{eqnarray} \label{#1}}
\newcommand{\eeqa}{\end{eqnarray}}
\newcommand{\beqn}{\begin{eqnarray*}}
\newcommand{\eeqn}{\end{eqnarray*}}
\newcommand{\diag}{{\rm diag}}
\newcommand{\rf}[1]{(\ref{#1})}
\newcommand{\bnab}{{\bf \nabla}}
\newcommand{\half}{\frac{1}{2}}
\newcommand{\gvk}{\ae}
\newcommand{\gD}{\Delta}
\newcommand{\dta}{\dot{a}}
\newcommand{\dtb}{\dot{b}}
\newcommand{\prt}{\partial}
\newcommand{\ddta}{\ddot{a}}
\newcommand{\gs}{\sigma}
\newcommand{\const}{{\rm const}}
\newcommand{\Gg}{\gamma}
\newcommand{\Dt}{\frac{d}{d{t}}}
\newcommand{\gG}{\Gamma}
\newcommand{\ga}{\alpha}
\newcommand{\gb}{\beta}
\newcommand{\dx}[2]{\frac{\partial{#2}}{\partial{x^{#1}}}}
\newcommand{\bF}{{\bf F}}
\newcommand{\bu}{{\bf u}}

\begin{document}


\begin{center}
{\Large \bf Effect of latent mass in inhomogeneous cosmological model with
perfect fluid and self-acting scalar field.} \\

\medskip
{V.M.~Zhuravlev$^1$, D.A.~Kornilov$^2$} \\
{Ulyanovsk State University \\
Institute for Theoretical Physics\\
$^1$ zhuravl@sv.uven.ru \\
$^2$ kda@sv.uven.ru}
\end{center}

\bigskip

{{\bf Abstract.} The inhomogeneous cosmological model with matter in the form
self-acting scalar field and perfect fluid is considered. On the basis
of exact solutions is considered the evolution of density
distribution of a matter in space on a background cosmological
expansion by the Universe. Is shown, the first, in such model the
equation of a matter state is variable in time and is closely
connected to character cosmological expansions. Secondly, it is
shown with point of view of the observer the Universe looks as
space flat, but with effect of latent mass. This effect consists
in that the mass of a perfect fluid by dynamic measurements
surpasses the own mass perfect fluid that is explained by presence
scalar field.
}
\bigskip

PACS-94: 04.20.-q
\bigskip

The usual approach to study cosmological models consists from two
stages. The first stage consists in construction of global
homogeneous and isotropic model of the Universe of FRW-type with a
scalar field and (or) matter in the form perfect fluid. At the
second stage it is studied evolution perturbation of matter
density on a background of the homogeneous metrics constructed at
the first stage. Thus the inhomogeneity of the Universe is taken
into account only in the first order of the perturbation theory,
that is connected with linearization of the dynamics equations.
However, finding out structure of the Universe on the large scales
(voids, strings) show that the process of their formation can be
essentially non-linear and for simulation of their occurrence it
is necessary to use some other approaches. In the present work the
attempt is undertaken to investigate properties of the self-
coordinated model of a special scalar field and perfect fluid
inducing a gravitational field on the basis of the exact solutions
of the Einstein equations, describing Universe with cosmological
evolution. In this case the distribution of matter in the Universe
are described exact, instead of by the approximate solutions of
the equations. Therefore in such model it is possible to study all
effects of interaction of three fields in an exact form in
formation various spatial structures of a matter in space.

The most suitable material for construction this models is well-known
in the General Relativity theory the Majumdar-Papapetrou
metrics \cite{TSEE,PM} originally used for construction self-gravitating
electrostatic or magnitoststic fields.

Basis of
suggested in the present work models is the generalization of the
Majumdar-Papapetrou metrics to describe not static effects
connected with cosmological expansion. Thus an electrostatic or
magnitoststic field is replaced by some generalized scalar field.
Nevertheless this scalar field has a similar energy-momentum
tensor to electrostatic field distinguished from them only by
presence of its self-action. Self-action of this field describes
by potential function which one in many respects determines
cosmological properties of model. The second kind of material
field in this model is the perfect fluid. In result the researched
model represents inhomogeneous cosmological model with self-acting
scalar field and matter with state equation generally varies in
due course. Last fact is important from the point of view modern
representations about existence of various epochs in evolution of
the Universe during the substance filling the Universe had various
properties. At an inflationary stage it is a matter close to
quasi-vacuum state, i.e. it prevails of a field component, in the
subsequent epoch it is the isotropic radiation ($p =\veps/3 $) and
now it is a matter close to a dust ($p\sim 0 $). In the present
work the attempt is done(made) to analyse such inhomogeneous
models from the point of view of evolution including all these
stages of development by the Universe.


1. General view of the metrics researched in the present work is
following
\beqa{Mabt}
&&ds^2=e^{A (x, y, z) +b (t)} dt^2-e^{-A (x, y, z) +a (t)}
\left (dx^2+dy^2+dz^2\right), \\
 \nonumber
&&g_{ik} = \diag\{e^{A+b}, -e^{-A+a}, -e^{-A+a}, -e^{-A+a} \},
   ~~ i, k=0,1,2,3,
\eeqa
where $A=A (x, y, z) $ - function of coordinates $x^1=x, x^2=y,
x^3=z $ and not dependent from $x^0=t $, and $a (t), b (t) $ -
some functions of time. The metric of such form with $a(t)=0,
b(t)=0$ is known in the classical theory of GR as the Majumdar-Papapetrou
metrics. In a case $a (t) \not=0, b (t) \not=0 $ these
not static metrics also describe some global cosmological dynamics
of the Universe with local inhomogeneity of the space it was
connected with function $A $. There is one more important
underclass of the metrics of this type having the following form
\beqa{Mabz}
&& ds^2=e^{A (x, y, t) +a (z)} \left (dt^2-dx^2-dy^2\right) -
e^{-A (x, y, t) +b (z)} dz^2.
\eeqa
It describes non-static gravitational processes in space with coordinate
$z$. Dependence from $z $ of functions $a, b $ in this
case is connected to some inhomogeneous and anisotropy space
lengthways the select axis $z$.

Let's consider the matter inducing metric property space - time of
the kind \rf{Mabt} the mix from perfect fluid with energy-momentum
tensor (TEM)
\beq{Tm}
  T^{(m)0}_0=\veps(x, y, z, t), ~T^{(m)1}_1=T^{(m)2}_2=T^{(m)3}_3=-
p(x,y,z,t), ~
     T^{(m)i}_k=0, ~i\not=k
\eeq
($\veps$ and $p$ - density of energy and pressure of a fluid) and
scalar field with energy-momentum tensor
\beq {Tphi}
      T^{(\phi)}_{ik} = -\bnab_i\phi\bnab_k\phi +
      \half g_{ik} g^{lm} \bnab_l\phi\bnab_m\phi + g_{ik} V (\phi,t)
\eeq
( $V (\phi, t) $ - potential of self-action of a scalar field). As
follows from \rf {Tphi} TEM of a scalar field considered in this
work differs from TEM of usual scalar field, for example Higg's
scalar field, by opposites sign, but coincides with sign of TEM of
an electrostatic field with potential $\phi$. However, equation
\rf{Tphi} differs from TEM of an electrostatic field by presence
of self-action potential.

The Einstein equations
\beq {EqEn0}
      G_{ik} = \gvk T_{ik}
\eeq
( $ \gvk=8\pi G/c^4 $ - Einstein's gravitational constant) for the
metrics \rf{Mabt} with TEM
$$
      T_{ik} =T^{(\phi)}_{ik} +g_{ij} T^{(m) j}_k
$$
are reduced to simple set from two equations, which is possible to
write down in the following form
\beqa{EqEn1}
&&p(x,y,z,t)=V(\phi)+g(t)e^{-b-A},\\
\label{EqEn2}
&&\veps(x,y,z,t)=c^2\rho(x,y,z,t)=\frac{1}{\gvk}e^{A-a}\gD A-V(\phi)
+\frac{\dta^2}{\gvk}e^{-b-A},
\eeqa
где
$$
g(t)=\frac{1}{\ae}\left[\frac{1}{2}\dta(\dtb-\dta)-\ddot{a}\right],
$$
and one statement
\beq {Aphi}
      \phi (x, y, z, t) = \frac {a (t) -A (x, y, z)} {\sqrt
{2\gvk}},
\eeq
identifying a scalar field with characteristic of the metrics.
To these equations it is necessary obviously to add the equation
arising the ambassador variations of a Lagrangian density of a
matter by $ \phi $. This equation has the following form
\beq {Eqphi}
     \gD A = \sqrt{2\gvk}e^{-A+a}\frac{\prt V(\phi)}{\prt \phi}+
     \left(\frac{\dtb-3\dta}{2}\dta -\ddta \right)e^{a-b-2A}
\eeq

Further, for function $A$ the equation \rf{Eqphi} should not on
the right contain obvious dependence from coordinate $t $, as
function $A $ is obvious from $t $ does not depend. It imposes
specific conditions on potential $V(\phi, t)$. By the elementary
kind $V (\phi, t) $ it is possible to satisfy to condition of
independence of the right part in \rf{Eqphi} from $t$, is
potential of a kind
\beq {V0}
      V (\phi) = V_0 \exp\{\sqrt{2\gvk} \phi \}.
\eeq
In this case we have
\beqa {EqEnab1}
&&\gD A = \gs e ^ {-2A}, \\
\label {Eqpr}
&& p=  g(t) e^{-b-A}+V(\phi)=p_0(t)e^{-A},\\
&& \veps=c^2\rho= g(t) e^{-b-A}-V(\phi)+\sqrt{\frac{2}{\gvk}}\frac{\prt V}{\prt \phi}
=p_0(t)e^{-A},
\eeqa
where
\beqn
&&p_0(t)= g(t)e^{-b}+V_0e^a,\\
&&\gs=2\gvk V_0e^{2a}+\left(\frac{\dtb-3\dta}{2}\dta- \ddta\right)e^{a-b}=\const.
\eeqn
Last statement is equation connecting $a$ and $b$, at any meanings
of parameters $\gs$ and $V_0$. From here for \rf{V0} we come to
the extreme rigid state matter equation
\beq {EqStM0}
   p = \veps,
\eeq
or to absence of a matter: at $p=0$ is automatically received
$\veps=0$.

That the state matter equation would have more general view, for
example, $p =\Gg(t)\veps$, it is necessary to require performance
of the following equation for $V (\phi, t) $:
$$
  V(\phi, t) +q(t) \exp\{\sqrt{2\gvk}\phi\} =\Gg(t)\left(-V(\phi,
t)+\sqrt{\frac{2}{\gvk}}\frac{\prt V}{\prt\phi}+
q(t)\exp\{\sqrt{2\gvk} \phi \}\right).
$$
Here
$$
q (t) =  g (t) e ^ {-b(t)-a(t)}.
$$
The general solution of this equation rather $V(\phi, t)$ at
parametrical dependence from $t$ has the following form
\beq{EqPhi}
   V (\phi, t) =V_1 (t) \exp\left\{ \frac{\Gg+1}{2\Gg} \sqrt{2\gvk}
   \phi\right\} -
   q (t) \exp\{\sqrt {2\gvk} \phi\},
\eeq
where $V_1(t)$ - any function $t$. The parameter $\Gg$ can be thus
function of time, and can and to not be. However in all cases,
when $ \Gg\not=1 $ and $ \Gg\not=0 $ the self-action potential
will be obvious function of time. For example, for the case, when
the substance represents by itself isotropic radiation with
$\Gg=1/3$ potential looks like
\beq{EqPhi1}
    V (\phi, t) = V_1(t) \exp\{2\sqrt {2\gvk} \phi \} -
   q (t) \exp\{\sqrt {2\gvk} \phi \}.
\eeq
In general case equation for $A (x, y, z) $ will look like
\beq{DA}
  \gD A=A_1e^{-(1+3\gamma)A/(2\gamma)}+A_2e^{-2A}.
\eeq

 Beacause the right member of the equation \rf{DA} did not contain a time $t$
 dependence, it is necessary that the functions $a,~b,~V_1$ satisfied to
 following requirements:
\beq{A1}
  A_1=\const=4\ae V_1(t) e^{3a},
\eeq
\beq{A2}
  A_2=\const=- \left[ \frac{\dta(\dtb+\dta)}{2}-\ddta\right] e^{a-b}.
\eeq

At once from \rf{A1} it is possible to obtain a kind of function
$V_1(t)$:
\beq{V01}
  V_1(t)=\frac{A_1\gamma}{\ae(1+\gamma)}e^{(1+3\gamma)a/(2\gamma)}.
\eeq

Equation \rf{A2} contains two unknown functions, one of them
$(b)$ remains uncertain and connected with selection of time variable.
For simplicity let's assume $b(t)=0$. It is means that variable $t$ is the
phisical time. In this case differential equation
$$
\ddot a-\frac{\dot a^2}{2}=A_2e^{-a},
$$
defines form of function $a(t)$, and it has the solution in an explicit
form in only elliptic functions of the first kind \cite{IntRad}.

Last equation determines evolution of the scale factor
$R(t)=\exp{\{-a(t)/2\}}$ too. In figure 1 the results of a numerical
analysis of the differential equation
\beq{R}
\frac{d^2 R}{dt^2}=-\frac{A_2}{2} R^3,
\eeq
that defines evolution of the scale factor.

\begin{figure}
 \psfig{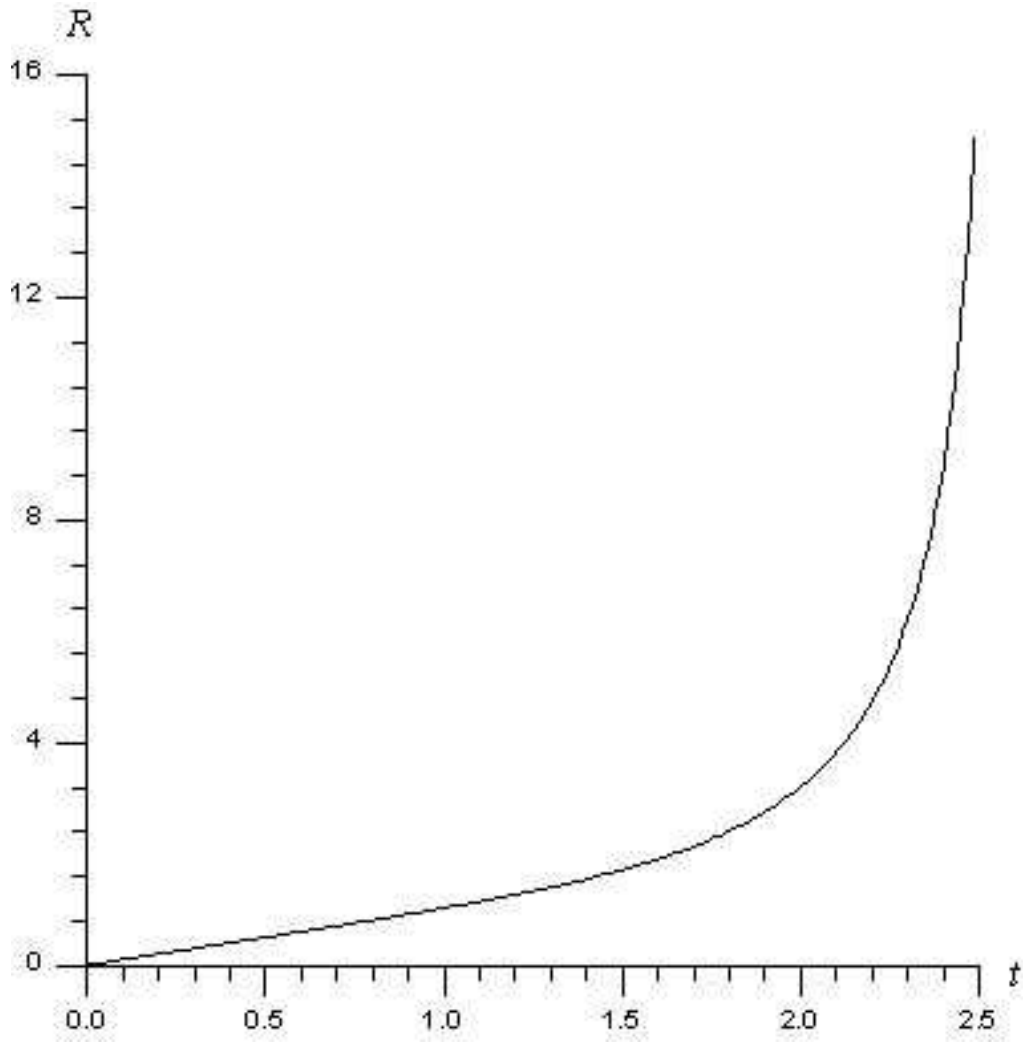}
\caption{{\it Evolution of the scale factor in case
 $A_2=-1$.
 Initial condition is follows:
 $R(0)=0,~\left. \frac{dR}{dt}\right|_{t=0}=1$.}}
\end{figure}

From the equation \rf{R} it is visible that the changes of evolution of the
scale factor are determined by the sign of the constant $A_2$. In
particular, in a case $A_2<0$, the second derivative on time
from the scale factor is positive (positive acceleration) and the expansion
of the Universe will be accelerated. For more best estimate
of velocity of expansion of the Universe in a fig. 1 the diagrams power
and exponential functions are given. It is visible, that the expansion at
a small times is a very good aproximated by a linear function, then
the very fast growth of the scale factor follows.

2. To find out properties of a field $\phi$ and its role in
represented theory it is important to study dynamics of a
trial particle in a gravitational field described by the metrics
\rf {Mabt}. Let's assume, that the field $\phi$ does not interact
directly with a usual matter. Then at absence of direct
interaction trial particle with a field $ \phi $ the equations of
its dynamics will look like
Geodetic
\beq {Equ}
   \Dt u^i +\gG^i_{kj} u^ku^j=0, ~~\frac {dx^k} {dt} =u^k,~i,j, k=0,1,2,3,
\eeq
where $u^k $ - 4-speed of a particle and $G^i_{kj}$ is the
Christoffel simbols. For the metrics \rf{Mabt} 4-speed of a
particle are normed by a condition
\beq {NMabt}
u^iu_i=g _ {ik} u^iu^k=e ^ {A} \left (e ^ {-2A} (u^0) ^2- (u^1)
^2- (u^2) ^2- (u^3) ^2\right) =c^2.
\eeq
where $c $ - speed of light.

Substituting in \rf{Equ} Christoffel symbols for the metrics
\rf{Mabt} we receive
\beqa {EquVIa}
&&\Dt u ^\ga = -\frac {1}{2}
\left (e^{2A} (u^0)^2 + (u^1)^2 + (u^2)^2 + (u^3)^2\right) A_{,\ga}+u^{\ga} (u ^\gb A_{,\gb}), \\
 \nonumber
&&\ga, \gb=0,1,2; \\
 \label {EquVI3}
&&\Dt u^0 =-u^{0} (u^\gb A_{,\gb}).
\eeqa

Having copied these equations for covariant components of 4-speed
of a particle using statement
$$
   u^0=e^{-A} u_0, ~~ u^1 =-e^{A} u_1, ~~ u^2 =-e^{A} u_2, ~~ u^3
=-e^{A} u_3,
$$
we receive
\beqa {EquNI0}
 &&\Dt u_0 = 0, \\
 \label {EquNIa}
 &&\Dt u_\ga = -\frac{1}{2} \left({2 (u_0) ^2} -c^2e ^ {A}
\right) A _ {,\ga}, ~~~\ga=0,1,2.
\eeqa
Thus covariant component $u_0 $ of 4-speeds of a particle is
constant and last three equations are equivalent to equations of
movement of a particle in a Newtonian field of gravitation:
\beq {CoVEq}
    \Dt u_\ga = -\Psi _ {,\ga} ~~~\ga=0,1,2,
\eeq
with gravitation potential
\beq {Psi}
    \Psi = {(u_0)^2} A-\frac {c^2} {2} e^{A}.
\eeq
Actually first member in this expression proportional to a square
of kinetic energy of a trial particle is the item connected with
potential forces of inertia and only second can be interpreted as
Newtonian potential of a field of gravitation. For interpretation
of movement the particles in this case are necessary for
considering the equations \rf {EquVIa}. This equations is possible
to write down as
\beq {EqDt}
    \Dt u^\ga =-\dx {\ga} \Phi+F^\ga,
\eeq
where $ \Phi $ now true potential of a field of gravitation:
\beq {Phi}
     \Phi = \frac {c^2} {2} e^{A} + \Phi_0,
\eeq
$ \Phi_0 $-any constant, and $ \bF $ is the gyrotropic force of
inertia arising at the expense of local rotation of reference
system. It will easily be convinced as $ \bF $ looks like
$$
      \bF = [\bu \times [\bu \times \bnab A]],
$$
where on the right there is a double vector product on flat
contravariant card of co-ordinates. As the force $ \bF $ is
gyrotropic, it does not make work.

Repeating the calculations have been carried out above for the
metrics \rf {Mabt} it is possible to receive similar to
\rf{EqDt} equations for contravariant and covariant component of
4-speeds in the metrics \rf{Mabz}.

3. We shall consider now the distribution of a matter corresponds
to potential fields of gravitation \rf{Phi}. Using \rf{EqEn2},\rf{V0}
we come to the following equation:
\beq{Phit}
\gD\Phi\equiv \frac{c^2}{2}e^{A}\left((\bnab A)^2+\gD A\right)
\eeq
Last member in the right part, square-law on a gradient of a
field, will be equal to energy ensity $ \veps_\phi $ of a scalar field
$ \phi $. Using \rf{EqEn2} the equation \rf{Phit} it is possible to write
down in a standard form
$$
\gD\Phi = 4\pi G \varrho,
$$
where
$$
\varrho= \rho_d+\frac{c^2}{2}V(\phi)+c^2(\bnab \phi)^2 e^{-\sqrt{2\ae}\phi}
$$
is complete density of all kinds of a matter in space: $ \rho_d $ -
density usual matter, and the other items represent energy density
 of a field $ \phi $  at the expense of self-action of a field $ \phi $.

Let's notice this result describe the presence of effect of latent mass.
Scalar field in this model not detected by not dynamic measurements, as is
actually connected by virtue of equality \rf{Aphi} with the metrics of
space - time and itself does not interract with usual matter. But at dynamic
measurements will be present additional mass. For any volume $V$ of space the
additional mass $M_l$ is equal to following
$$
M_l=\int\limits_\Omega
\left[ \frac{c^2}{2}V(\phi)+c^2(\bnab \phi)^2 e^{-\sqrt{2\ae}\phi}
\right] d\Omega.
$$
This formula show that in considarating model the effect of latent mass exists.
The value of latent mass define by function of $A(x,y,z)$ and dependent from
cosmological variation of scale factor that equail to $R(t)=\exp{-a(t)/2}$.

{\it Acknowledgements}. Work is supported by Russian Fund of Basic Researches
(grant 01-98-18040).

\end {document}